\documentclass[pra,twocolumn,showpacs,floatfix]{revtex4}
\usepackage{amsmath}
\usepackage{amsfonts}
\usepackage{graphicx}
\usepackage[usenames]{color}
\newcommand{\beq}{\begin{equation}}
\newcommand{\eeq}{\end{equation}}
\newcommand{\beqa}{\begin{eqnarray}}
\newcommand{\eeqa}{\end{eqnarray}}
\newcommand{\ba}{\begin{array}}
\newcommand{\ea}{\end{array}}

\begin{document}

\title{Self-trapping of a Fermi super-fluid
in a double-well potential in the BEC-unitarity crossover}
\author{S. K. Adhikari$^{1}$\footnote{adhikari@ift.unesp.br;
URL: www.ift.unesp.br/users/adhikari}, Hong Lu$^2$ and Han Pu$^{1,2} $\footnote{hpu@rice.edu}} \affiliation{$^1$Instituto de
F\'{\i}sica Te\'orica, UNESP - S\~ao Paulo
State University,
01.140-070 S\~ao Paulo, S\~ao Paulo, Brazil\\
$^2$Department of Physics and Astronomy, and Rice Quantum Institute,
Rice University, Houston, TX 77251, USA}

\begin{abstract}
We derive a generalized Gross-Pitaevskii
density-functional equation appropriate to study the Bose-Einstein
condensate (BEC) of dimers formed of singlet spin-half Fermi pairs in the
BEC-unitarity crossover 
while the dimer-dimer scattering length $a$
changes from 0 to $\infty$. Using an effective one-dimensional form
of this equation, we study the phenomenon of dynamical self-trapping
of a cigar-shaped Fermi super-fluid in the entire BEC-unitarity
crossover {{in a double-well potential}}. A simple two-mode model 
is constructed to provide
analytical insights. {{We also discuss the consequence of our
study on the self-trapping of an atomic BEC in a
double-well potential.}}

\end{abstract}


\maketitle

\section{Introduction}

\label{I}

{After the experimental realization of Bose-Einstein condensate
(BEC) and its controlled study under different trapping conditions \cite{book},
there have been many interesting experiments with
a cigar-shaped BEC
in a quasi one-dimensional (1D) trap with a tight transverse
confinement \cite{gor}. Along the axial direction several different types of
traps have been employed: harmonic \cite{book}, 
double-well \cite{GatiJPB2007},  periodic
optical-lattice \cite{kast}, and bichromatic optical-lattice \cite{bich} 
traps.  Many novel phenomena have been
predicted and observed in such quasi-1D setting. Of these, the ones
worth mentioning include the formation of bright \cite{bright}, gap
\cite{gap}, and dark \cite{dark} solitons, self-trapping
\cite{trappingPRL,GatiJPB2007} and Josephson oscillation
\cite{GatiJPB2007,cataliotti,joseBEC}.

Macroscopic dynamical self-trapping and Josephson oscillation were
predicted theoretically
\cite{java,trappingPRL,Gong,OstrovskayaPRA2000,TS} and observed
experimentally \cite{cataliotti,GatiJPB2007}. Josephson effect was
observed in super-fluid (SF) $^3$He  \cite{He3} and $^4$He
\cite{He4}. After the experimental observation of BEC in a 
optical-lattice trap
\cite{cataliotti}, controlled studies of Josephson oscillation and
self-trapping in a cigar-shaped  BEC seems well under control
\cite{GatiJPB2007}. The studies of such phenomena in a cigar-shaped
BEC usually employ a double-well potential \cite{GatiJPB2007}.
In the simplest case of
such a symmetric 1D potential with the origin of the axial
coordinate $x$ set at the trap center, under certain initial
conditions, when a BEC is released with a population imbalance
between two sides of $x=0$, it executes undamped Josephson
oscillation on both sides of the trap center maintaining a
time-averaged population imbalance equal to zero. Under different
initial conditions, the BEC exhibits self-trapping, occupying
preferably one side of the trap, thus maintaining a definite
non-zero value of time-averaged population imbalance. The
understanding of the transition from Josephson oscillation to
self-trapping and vice versa has been the topic of many recent
investigations \cite{GatiJPB2007}.

A SF Fermi gas in a double-well potential is perhaps even more
interesting, nevertheless much less studied \cite{sala}.
(There have been studies of Josephson oscillation of a Fermi gas
in a OL potential \cite{ferop}).
Such a
trapped SF Fermi gas gives us the unique opportunity to study the
Bardeen-Cooper-Schrieffer (BCS) to BEC { crossover} in a
two-component Fermi gas under an entirely different set-up. The
BCS-{BEC crossover} can be realized by varying the
attraction between the spin-half fermions forming pairs using
the Feshbach resonance technique. As the attraction is increased
from zero, the simple BCS SF turns into a complex
Cooper-pair-induced strongly interacting SF and at unitarity (when
the Fermi-Fermi scattering length $a_f$ tends to infinity), it is
possible for the Cooper pairs to turn spontaneously into Fermi
dimers (two-body bound state of fermionic atoms) and the BCS SF turns
into a BEC of dimers.

After the experimental realization \cite{exp23} of the
BCS-{BEC crossover} in a trapped Fermi SF by varying the
atomic interaction near a Feshbach resonance, there have been
renewed interests \cite{exp10,DF2,rmp2} in the study of a Fermi SF at
unitarity and beyond in the BEC region where we have the BEC of
dimers. One can thus recover the bosonic behavior in the BEC limit
of the crossover (when the dimer-dimer scattering length $a$ tends
to zero), while expecting new and distinct behavior in the vicinity
of the unitarity regime. Moreover, on the experimental front it is
easier to realize a controlled BEC-unitarity crossover
(BEC side of the BCS-BEC crossover) of the Fermi
SF than the BCS-unitarity crossover 
(BCS side of the BCS-BEC crossover), as the super-fluid transition
temperature in the BCS side of the crossover is very low and
difficult to achieve.

Here we present a unified Galilean-invariant dynamical equation for the
study of the BEC-unitarity crossover of a cigar-shaped BEC of dimers
formed of Fermi atoms. In the BEC limit of small dimer-dimer scattering
length $a,$ the present equation reduces to the usual Gross-Pitaevskii
(GP) equation \cite{GP} for bosons, and in the unitarity limit it yields
a density-functional (DF) equation \cite{DFT} for fermions. Hence we
call this equation a DF GP equation for a Fermi SF valid in the
BEC-unitarity crossover. For the study of a cigar-shaped Fermi SF along
the BEC-unitarity crossover, we reduce the present DF GP equation to a
quasi-1D form {{ and use this reduced equation to the study of the
self-trapping of Fermi SF in a double-well potential}}.
This analytic development is presented in Sec. \ref{II}.
{{This reduced equation also describes an atomic BEC with
repulsive atomic interaction in the BEC-unitarity crossover with
different numerical value(s) of certain parameter(s), and hence the
results of the present investigation are also applicable to the self-trapping
of a repulsive BEC in a double-well potential.}}

The numerical simulation with the time-dependent quasi-1D equation
for a cigar-shaped Fermi SF in a symmetric double-well potential
with an initial population imbalance between two wells reveals
interesting features of  the Josephson oscillation and self-trapping
across the BEC-unitarity crossover. In the limit of zero
nonlinearity one has AC Josephson oscillation. As nonlinearity is
increased by increasing the dimer-dimer scattering length or the
number of particle, the Josephson oscillation stops and self-trapping
emerges {{for a double-well potential with appropriate parameters}}. 
With 
further increase of  nonlinearity 
self-trapping is destroyed 
and the population in the two wells executes irregular
oscillation. {For very large nonlinearity, however, the regular
Josephson oscillation comes back. Nevertheless, 
for a small number of particles, 
the critical nonlinearity required for one of these phenomenon may not be
attained and that particular phenomenon may not be realized.
(The nonlinearity actually saturates for a large value of $a$ and hence cannot 
be arbitrarily increased by increasing $a$ as one approaches unitarity for a small 
number of atoms.)}
These features are discussed in detail in Sec.
\ref{III} where we present the numerical results.

In Sec. \ref{IV}  we present a simple analytic two-mode model to
understand the essential features of the numerical results reported
in Sec. \ref{III} and also point out the limitation of the two-mode
model. Finally, in Sec. \ref{V} we present a brief summary and
conclusion of the present investigation.

\section{DF GP equation
for a Fermi SF in the BEC-unitarity crossover}

\label{II}

At unitarity the following density-functional (DF) equation for trapped SF
fermions \cite{AS2,AS3}
has produced results for energy in close agreement
with independent Monte-Carlo calculations \cite{MC}
\begin{eqnarray} \label{eq1}
\left[-\frac{\hbar^2}{2m}\nabla^2+U+\mu(n)-i\frac{\partial}{\partial
t}\right]\Psi({\bf r},t)=0, \end{eqnarray}
where $U$ is the trapping
potential, $m$
is the mass of a dimer (twice the atomic mass),
$\mu(n)=\xi\hbar^2n^{2/3}/m$ is the bulk chemical potential
of dimers with density (of dimers)  $n=|\Psi|^2$,
and
$\xi=2(6\pi^2)^{2/3}\zeta$.
The normalization condition of the DF wave function
is
$\int|\Psi|^2 d^3r=N$, where $N$
is the number of dimers.
At unitarity the only length scale is
$n^{-1/3}$, and from dimensional argument the chemical potential $-$ and
all energies of the trapped SF fermions
$-$ have the above universal
form \cite{uni}.

There have been many theoretical \cite{the1,the2} and experimental \cite{exp} investigations
which
extracted the value of the constant $\zeta$ for fermions, and the most
accurate value of this constant is
given by independent Monte-Carlo calculations by two groups \cite{the1}:
$\zeta
\approx 0.44$, consequently, $\xi \approx
13.37$ and we shall use this value of $\xi$ in the present study.
{{
For a
trapped atomic BEC, the energy and chemical potential have the same
universal form: $\sim \xi \hbar^2 n^{2/3}/m$ \cite{cow}, where now
mass $m$ and density $n$ refer to bosons.  For fundamental bosonic
atoms, microscopic numerical calculation based on Jastrow
variational wave function yielded for the constant $\xi$ a slightly
different value $\xi=22.22$ \cite{cow}.}  With this modification in the value 
of $\xi$, the present
investigation could be applied to the study of self-trapping of an atomic BEC.}

At unitarity the Fermi pair can stay as a Cooper pair or a dimer and
transform into each other without transfer of energy and Eq. (\ref{eq1})
can describe both the Cooper pair and dimer phases. Here we interpret
Eq. (\ref{eq1}) as the equation for  dimers.  At unitarity the
scattering length $a$ of two dimers goes to infinity: $a\to \infty$.
(Actually, at unitarity, the scattering length of two Fermi atoms $a_f
\to \infty$. Model studies have
indicated that $a\propto a_f$ \cite{petrov}. Consequently,
at unitarity we take $a\to
\infty$.)

Although Eq. (\ref{eq1}) describes both the dimer SF and the
Cooper-pair induced BCS SF at unitarity, the bulk chemical potential
$\mu(n)$ appearing in this equation should be interpreted
differently. For the BCS SF it originates from the kinetic energy of
Fermi atoms put in different quantum orbitals consistent with the
Pauli principle discounted for by the negative attractive energy due
to atomic interaction. For the dimer SF it originates solely from
the repulsive interaction energy among dimers. In the BCS limit as
$a_f \to 0^-$ we have the finite nonlinear term
$\mu(n)=2E_F=2(6\pi^2)^{2/3}\hbar^2n^{2/3}/m$ \cite{AS2} in
Eq.~(\ref{eq1}) originating from the kinetic energy of Fermi atoms
with negligible contribution from inter-atomic attraction. On the
other hand, in the BEC limit as $a \to 0^+$ the nonlinear term for
dimers reduces to zero and at unitarity the nonlinear term in
Eq.~(\ref{eq1}) originates from the saturation of repulsive
dimer-dimer interaction as $a\to \infty$.

For the Fermi SF of dimers ({{and also for an atomic BEC}}) in the BEC-unitarity crossover the following
two leading terms of the bulk chemical potential of a dilute uniform gas
can be obtained \cite{nilsen}
from the expression for energy per particle as obtained
by Lee, Huang, and Yang \cite{lee}
\begin{eqnarray} \mu(n,a)=({4\pi\hbar^2 a n}/{m})
\left(1+\alpha (n^{1/3}a)^{3/2}+...\right),\label{exp}
\end{eqnarray}
where $\alpha= 32/(3\sqrt \pi)$ and
$n^{1/3}a$ is the dimensionless gas parameter. In
this expression the scattering length $a$ must be positive ($a>0$)
corresponding to a repulsive interaction. Higher-order terms of
expansion (\ref{exp}) has also been considered \cite{bra}; the
lowest order term was derived by Lenz \cite{lenz}. Considering only
the lowest-order term in expansion (\ref{exp}), appropriate in the
BEC limit as $a,a_f\to 0^+$, the dimers obey the usual
GP equation \cite{GP}
\begin{eqnarray} \label{eq1b}
\biggr[-\frac{\hbar^2}{2m}\nabla^2+U+\frac{4\pi\hbar^2a}{m}|\Psi|^2
-i\frac{\partial}{\partial t}\biggr]\Psi({\bf r},t)=0
\,.\end{eqnarray}

Considering the second term in the expansion (\ref{exp}), in the
BEC limit, the following modified GP equation for dimers can be
written following the suggestion of Fabrocini and Polls \cite{polls}
\begin{eqnarray} \label{eq2}
&&\biggr[-\frac{\hbar^2}{2m}\nabla^2+U+\frac{4\pi\hbar^2a}{m}|\Psi|^2
\left(1+\alpha a^{3/2}|\Psi| \right)\nonumber\\ &&
-i\frac{\partial}{\partial t}\biggr]\Psi({\bf r},t)=0
\,.\end{eqnarray} Equation (\ref{eq2}) provides an adequate
correction to the GP equation~(\ref{eq1b}) for small $a$. But as $a$
increases and diverges at unitarity, the nonlinear term should
saturate to the finite universal nonlinear term $\mu(n)$ of Eq.
(\ref{eq1}) and not diverge like the nonlinear terms of the GP
equation (\ref{eq1b}) and of the Fabrocini-Polls equation
(\ref{eq2}). The chemical potential and energy should not diverge at
unitarity, as the interaction potential remains finite in this limit,
although the scattering length $a$ diverges. In the weak-coupling GP
limit, the scattering length serves as a faithful measure of
interaction. But as $a$ increases, it ceases to be a measure of
interaction. For a general scattering length, an exact expression of the chemical potential is not available. However, a recent quantum Monte Carlo study maps out the equation of state in the entire BEC-BCS crossover regime \cite{the1}.

Following a recent suggestion \cite{AS1},
for the full BEC-unitarity crossover we consider  the
DF GP equation for the dimer SF
providing a smooth interpolation between Eqs. (\ref{eq1})
and (\ref{eq2}):
\begin{eqnarray} \label{eq3}
\biggr[-\frac{\hbar^2\nabla^2}{2m}+U+\mu(n,a)
  -i\frac{\partial}{\partial t}\biggr]\Psi({\bf
r},t)=0\,,
\end{eqnarray}
\begin{eqnarray}
\mu(n,a)&=&\frac{4\pi\hbar^2a}{m}|\Psi|^2 \nonumber \\ 
&\times &
\left(\frac{1+\alpha(1+\delta)
a^{3/2}|\Psi|}{ 1+\alpha \delta a^{3/2}|\Psi|+\gamma (1+\delta)a ^{5/2}
|\Psi|^{5/3}} \right) \,,  \label{eq32}
\end{eqnarray}
where $n=|\Psi|^2$ and $\delta$ and $\gamma$ are yet unknown
constants satisfying $\gamma =4\pi\alpha/\xi$. 
({{These equations are also valid for an
atomic BEC  with $m$ and $a$ representing atomic mass and scattering
length, respectively, and with $\xi=22.22$ \cite{cow}.}})
By construction,   Eq. (\ref{eq32}) yields the limit 
(\ref{exp}) for small $a$;  it also has the correct behavior at unitarity. 
{{ Using a similar expression for 
$\mu(n,a)$ in the BCS-unitarity crossover \cite{ska}
the constant $\delta$ was
 calculated \cite{sadhan}
by requiring that the 
first derivative of $\mu(n,a)$ with respect to 
$(an^{1/3}$)
be continuous at 
unitarity. The condition for continuity yields a small value for $\delta
(\sim 0.04)$. } }
However,  we shall take $\delta =0$ in this study.
This  will make  further analytical development 
easier while maintaining the first derivative of $\mu(n,a)$ with respect to
$(an^{1/3}$)
approximately  continuous at
unitarity.
A
set of equations, similar to Eqs.~(\ref{eq3}) and (\ref{eq32}), for
fundamental bosons, and not for composite dimers, produced results
for energy \cite{AS2} of a trapped condensate in agreement with
Monte-Carlo calculations \cite{MCBOSE}. A similar equation for the
BCS-unitarity crossover  produced results for energy \cite{AS2,ska}
of a trapped BCS SF in agreement with Monte-Carlo calculations
\cite{MCBCS}. Different parametrization of the chemical potential
in the BEC-BCS crossover have been proposed in the literature
\cite{chempot}. We do not expect our results in this work will be
sensitive to which specific form for $\mu(n,a)$
we choose to use here. Furthermore,
it has been shown that the DF GP equation (\ref{eq3}) is equivalent
to the quantum hydrodynamic equations for dimers \cite{AS2,chempot}
\begin{eqnarray*}
\frac{\partial n}{\partial t}+\nabla \cdot (n {\bf v})=0 \,,\\
m \frac{\partial {\bf v}}{\partial
t}+\nabla\biggr[-\frac{\hbar^2}{2m} \frac{\nabla^2 \sqrt n}{\sqrt
n}+\frac{mv^2}{2}+U+ \mu(n,a)\biggr]=0\,,
\end{eqnarray*}
if we identify
\begin{eqnarray*}
\Psi({\bf r},t) &=& \sqrt{n({\bf r},t)}\, e^{is({\bf r},t)} \,,
\\
{\bf v} &=& \hbar\nabla s/m \,,
\end{eqnarray*}
where $s$ is a phase, and $\bf v$ is the velocity.

For a cigar-shaped SF, where the transverse trapping is very strong,
the interesting dynamics is confined in the axial direction and in
the transverse direction the system is confined in its ground state.
In such a quasi-1D geometry,  the axial and transverse coordinates
decouple and it is useful to write an effective 1D equation for the
dynamics of a cigar-shaped SF and we perform the same in the
following. For the cigar-shaped double-well trap,
\begin{eqnarray*}
U({\bf r})&=& V(x)+m \omega^2 (x^2+\lambda^2 z^2+\lambda^2 y^2)/2
\,,\\ V(x)&=& A\hbar\omega \exp(-\kappa m \omega x^2/\hbar)\,,
\end{eqnarray*}
where $\lambda \gg 1$, and $A$ and $\kappa$ are two dimensionless
parameters characterizing the strength and width of the barrier,
respectively, it is appropriate to take $\Psi({\bf
r},t)=\psi(x,t)\phi(y)\phi(z)$ with $\phi(y)= [m \omega
\lambda/(\hbar\pi)] ^{1/4} \exp[-m \omega\lambda y^2/(2\hbar)]$
{{representing the harmonic-oscillator ground state in the
transverse direction} and $\psi(x,t)$ representing the essential
dynamics in
the $x$ direction}. The
potential $V(x)$ together with the harmonic trap $m\omega^2 x^2/2$
simulate a double-well in the axial $x$ direction.
Multiplying Eqs. (\ref{eq1}) and (\ref{eq2}) by
$\phi(y)\phi(z)$ and integrating over $y$ and $z$ we get the following 1D
equations \cite{AS3,cigar}
\begin{eqnarray}\label{eq4}
\biggr[ -\frac{1}{2}\frac{\partial^2 }{\partial x^2}+
U(x)
+\xi \frac{3}{5}\left(\frac{\lambda n}{\pi}\right)^{2/3}
-i\frac{\partial}{\partial
t}\biggr]\psi(x,t)=0, \end{eqnarray}
\begin{eqnarray}\label{eq5}
\biggr[ -\frac{1}{2}\frac{\partial^2 }{\partial x^2}
+U(x)
+2a\lambda n \left(  1+\alpha a^{3/2}\frac{4}{5}\frac{\sqrt{n\lambda}}
{\sqrt\pi}
   \right)\nonumber \\
-i\frac{\partial}{\partial
t}\biggr]\psi(x,t)=0,
\end{eqnarray}
where ${{n=|\psi(x,t)|^2}}, U(x)=Ae^{-\kappa x^2}+
{x^2}/{2}
$
represents the double-well potential and
$n=|\psi|^2$ and
we use harmonic oscillator dimensionless units $\hbar=m=\omega=1$.
All lengths are now expressed in oscillator unit $\sqrt{\hbar/(m\omega)}$ and
time in $\omega^{-1}$, and $\psi$ is normalized as $\int_{-\infty}^{\infty }
dx |\psi(x,t)|^2=N$.

A simple DF GP equation interpolating between Eqs.~(\ref{eq4}) and
(\ref{eq5}) is
\begin{eqnarray}\label{eq6}
\biggr[ -\frac{1}{2}\frac{\partial^2 }{\partial x^2}
+U(x)+ \mu(n,a)
-i\frac{\partial}{\partial t}\biggr]\psi(x,t)=0\,, \\
\mu(n,a) = 2a\lambda n \left( \frac{1+\frac{4}{5}
\frac{\sqrt\lambda}
{\sqrt\pi} \alpha a^{3/2}
\sqrt n}{1+\beta a^{5/2}
n^{5/6}}
   \right)\,,\label{eq6x}
\end{eqnarray}
where $n=|\psi|^2$ and $\beta = 8\alpha \lambda^{5/6}\pi^{1/6}/(3\xi)$. Equation
(\ref{eq6}) reduces in the BEC $a\to 0^+ $ limit to Eq.~(\ref{eq5})
and in the unitarity $a\to \infty$ limit  to Eq.~(\ref{eq4}). We
shall use Eq.~(\ref{eq6}) for the description of self-trapping and
Josephson oscillation in the BEC-unitarity crossover of fermions.

\section{Two-Mode Model of the Fermi SF}

\label{IV}

\subsection{Two-Mode Model}

Before we present the full numerical results, it is instructive to 
consider the so-called two-mode model \cite{trappingPRL} which is widely 
used in the study of BEC in a double-well potential, and more recently, 
has been used in the investigation of Fermi SF across a weak link 
\cite{newref}. Due to its simplicity, the two-mode model can provide 
many useful insights. Here we construct the corresponding two-mode model 
of the Fermi SF based on Eq.~(\ref{eq6}).

To this end, we decompose $\psi(x,t)$ as
\begin{equation}
\psi(x,t) = \psi_1(t) \phi_1(x) + \psi_2(t) \phi_2(x) \,,
\label{dec}
\end{equation}
where the spatial mode functions $\phi_{1,2}(x)$ are assumed to be real,
satisfy the orthonormal condition
\[ \int \phi_i (x) \phi_j(x) dx =\delta_{ij} \,,\]
and are localized in
each of the two wells, respectively. $\psi_{1,2}(t)$ are in general
complex and satisfy the condition
{ $|\psi_{1,2}(t)|^2=N_{1,2}(t)$ so that }
\[ |\psi_1(t)|^2 + |\psi_2(t)|^2 = N_1(t)+N_2(t) =N \,.\] Inserting
the decomposition (\ref{dec}) into Eq.~(\ref{eq6}), integrating out
the spatial degrees of freedom, we obtain the following equations of
motion for $\psi_{1,2}(t)$:
\begin{eqnarray}
i \dot{\psi}_1 &=& E_1\, \psi_1 +{{ {\cal E}_{1}}}(|\psi_1|)\, \psi_1 -
{\cal K}\, \psi_2 \,,  \label{2modeA} \\
i \dot{\psi}_2 &=& E_2\, \psi_1 + {{ {\cal E}_{2}}}(|\psi_2|)\, \psi_2 -
{\cal K}\, \psi_1 \,,\label{2modeB}
\end{eqnarray}
where
\begin{eqnarray}
&&E_i = \int dx \, \phi_i(x) \left(-\frac{1}{2} \frac{d^2}{dx^2}+U(x) \right)
\phi_i (x) \,, \\
&&{{ {\cal E}_{i}}}(|\psi_i|) = \int dx\, \phi_i(x) \mu(n_i,a)  \phi_i(x) \,, \\
&&{\cal K} = - \int dx \, \phi_1(x) \left(-\frac{1}{2}
\frac{d^2}{dx^2}+U(x) \right) \phi_2(x) \, ,
\end{eqnarray}
with
$n_i = |\psi_i \phi_i|^2$.
Here we have neglected
integrals involving spatial overlaps of $\phi_1(x)$ and $\phi_2(x)$.

For simplicity,
we assume a symmetric double well with $U(x)=U(-x)$ so
that $\phi_1(x)=\phi_2(-x)$ and consequently, we have $E_1=E_2$ and
${{{\cal E}_{1}(|\psi|)={\cal E}_ {2}(|\psi|) 
\equiv {\cal E}(|\psi|)}}$. Let us write the waves $\psi_{1,2}$
in terms of its amplitude $\sqrt{N_{1,2}}$ and phase ($\theta_{1,2}$) 
\[ \psi_{1,2} = \sqrt{N_{1,2}}\, e^{i\theta_{1,2}} \,
\] and define a pair of conjugate variables: \[ S \equiv
(N_1-N_2)/N\,,\quad \theta \equiv \theta_2 - \theta_1 \,.\]
Here the variable $S$ denotes the population imbalance between the two 
wells and $\theta$ is the phase difference.
{{After some straight-forward algebra,  the following
equations of motion for $S$ and $\theta$ can be derived
from Eqs. (\ref{2modeA}) and (\ref{2modeB}) by equating the real
and imaginary parts of both sides: }}
\begin{eqnarray*}
\dot{S} &=& -{2{\cal K}}  \sqrt{1-S^2} \,\sin \theta \,,\\
\dot{\theta} &=&   \left[{\cal E}(\sqrt{N_1}) -{\cal E}(\sqrt{N_2})
\right] +  {2{\cal K}}  \frac{S}{\sqrt{1-S^2}} \,\cos \theta \,.
\end{eqnarray*}
These are the two-mode equations for the Fermi SF. Note that the
two-mode equations describing weakly-interacting bosons in
double-well potential \cite{trappingPRL} are recovered if we take
$\mu(n,a) = 2 a \lambda n$ and, correspondingly, ${\cal E}(\sqrt{N_i})
= 2a \lambda N_i \int |\phi_i|^4 dx$.

The two-mode equations can be cast into the canonical form \[
\dot{S} = -\frac{\partial H}{\partial \theta} \,,\quad \dot{\theta}
= \frac{\partial H}{\partial S} \,, \] with the classical
Hamiltonian defined as
\begin{equation}
H = \int \left[{\cal  E}(\sqrt{N_1}) -{\cal E}(\sqrt{N_2}) \right]\,dS
-2{\cal K}\sqrt{1-S^2} \cos \theta \, . \label{HC}
\end{equation}
By studying the properties of this Hamiltonian, we can tell whether
the system should exhibit self-trapping or Josephson oscillation.

\subsection{Fermi SF at Unitarity}
As a concrete example, let us consider the Fermi SF at unitarity
where $a \to \infty$ and
{{from Eq. (\ref{eq6x}) we find}}
\[ \mu(n,a) = \mu(n)= \frac{3\xi}{5}
\left(\frac{\lambda n}{\pi} \right)^{2/3} \,.\] It follows that \[
{\cal E}(\sqrt{N_i}) = UN_i^{2/3} \,,\] with $U =( {3\xi}/{5}) (
{\lambda}/{\pi} )^{2/3}
\int
|\phi_i|^{10/3} dx$. The classical Hamiltonian takes the form:
\[
\frac{H}{2{\cal K}} = \frac{3 \Lambda}{5} \left[ (1+S)^{5/3}+(1-S)^{5/3}
\right] -  \sqrt{1-S^2} \cos \theta \,.
\] where $\Lambda = (N/2)^{2/3}U/(2{\cal K})$
measures the ratio of the strength of the nonlinearity
$(N/2)^{2/3}U$ and the tunneling energy $2{\cal K}$.

\begin{figure}
\begin{center}
\includegraphics[width=.49\linewidth]{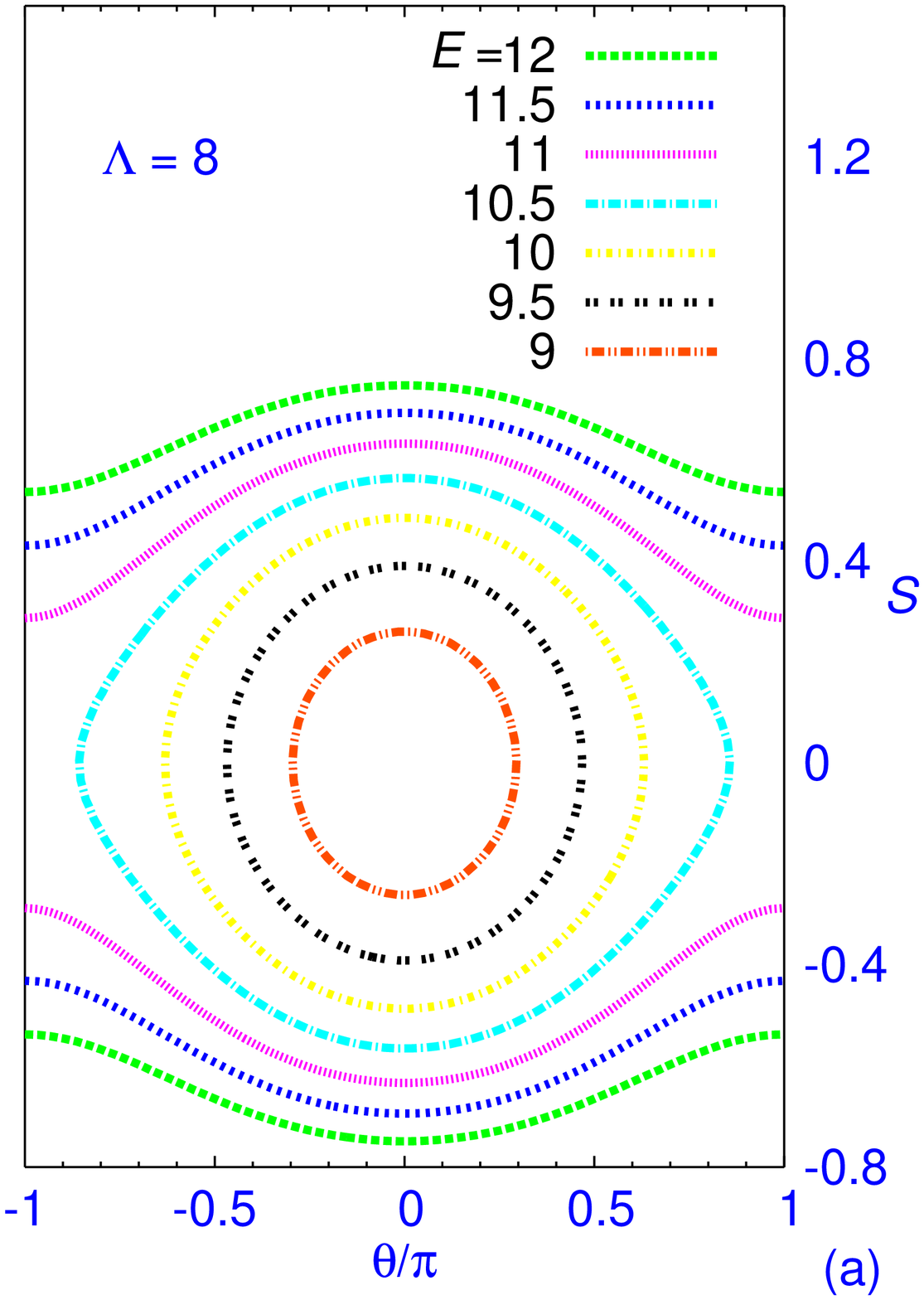}
\includegraphics[width=.49\linewidth]{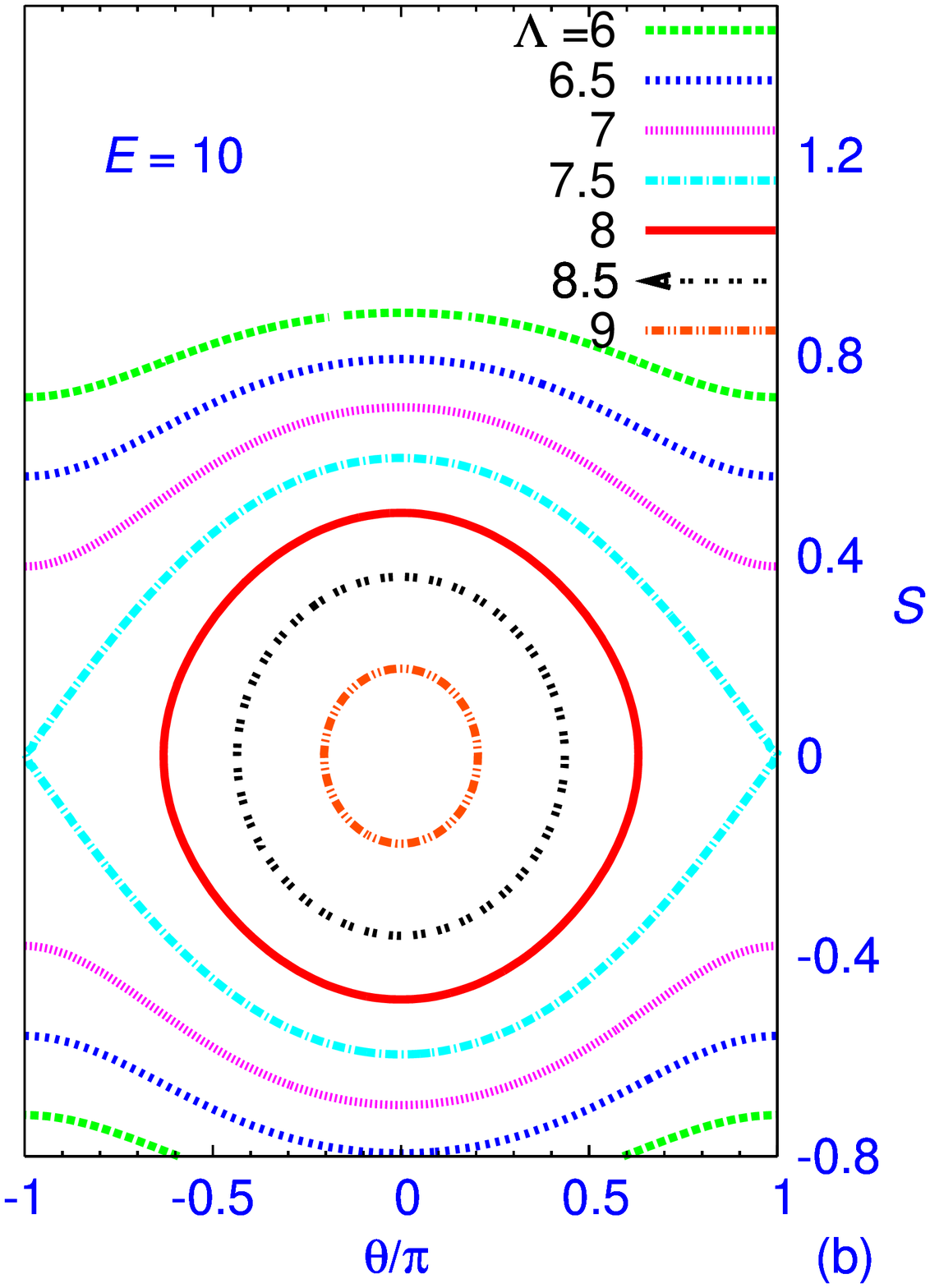}
\end{center}
\caption{(Color online){{Equi-energy}} phase contour plot of the unitary Fermi SF.
(a) Contours of different energies for $\Lambda=8$. (b) Contours of
$E=10$ for different values of $\Lambda$. Energy is in units of
$2{\cal K}$.} \label{fig3}
\end{figure}

As can be seen from Fig.~\ref{fig3}, if we draw equi-energy contours
of $H$ in the phase space of $(S,\theta)$, we can see two types of
contours: those that form closed loop and those do not. The division
of these two types of contours occurs at the critical energy (we
take $2{\cal K}$ as the units for energy) \[E_{\rm crit} =
H(S=0,\theta=\pi)=\frac{6\Lambda}{5}+1 \,.\] If the system has
energy $E<E_{\rm crit}$, its dynamics will follow the closed
contours, and both $S$ and $\theta$ will oscillate in time. In
particular, the population imbalance $S$ oscillates around 0 and the
time averaged population imbalance vanishes. This corresponds to the
AC Josephson regime. On the other hand, if the system has energy
$E>E_{\rm crit}$, it will follow the open contours where $\theta$
will grow indefinitely and $S$ will oscillate around a non-zero
value and will never cross the $S=0$ line. This corresponds to the
self-trapping regime.

Given the initial values $S(0)$ and $\theta(0)$, the system moves
on a contour of constant energy given by
\begin{eqnarray*}
E_0 &=& \frac{3\Lambda}{5} \left[ (1+S(0))^{5/3}+(1-S(0))^{5/3} \right] \\
&& \;\;
-\sqrt{1-S(0)^2} \cos \theta(0) \,.
\end{eqnarray*}
The condition for self-trapping \[E_0 > E_{\rm crit} =
\frac{6\Lambda}{5} +1 \,,\] may be recast into the form
\begin{equation}
\Lambda > \Lambda_c = \frac{5}{3} \frac{1+\sqrt{1-S(0)^2} \,\cos
\theta(0)}{(1+S(0))^{5/3}+(1-S(0))^{5/3} -2} \,. \label{Lambdac}
\end{equation}
In other words, within the two-mode model, the ratio of the
nonlinear strength and the tunneling energy, $\Lambda$, determines
whether the system should be self-trapped or not.

To determine the values of these quantities, we need to choose
properly the spatial mode functions $\phi_{1,2}(x)$. A reasonable
choice is given by \cite{trappingPRL,I2M}
\[ \phi_{1,2} (x)= \frac{\phi_+(x) \pm \phi_- (x)}{\sqrt{2}} \,,\]
where the normalized wave functions $\phi_{\pm}(x)$ are the
lowest-energy symmetric and antisymmetric stationary solutions to
the time-independent DF equation:
\[\bar  \mu_\pm \phi_{\pm} = \left[ -\frac{1}{2}\frac{d^2}{dx^2}
+U(x) +\frac{3\xi}{5} \left(\frac{N\lambda}{\pi}\right)^{2/3}
|\phi_{\pm}|^{4/3} \right]\,\phi_{\pm} \,,\]
with chemical potential $\mu_\pm$.

\begin{figure}
\centering
\includegraphics[width=\linewidth]{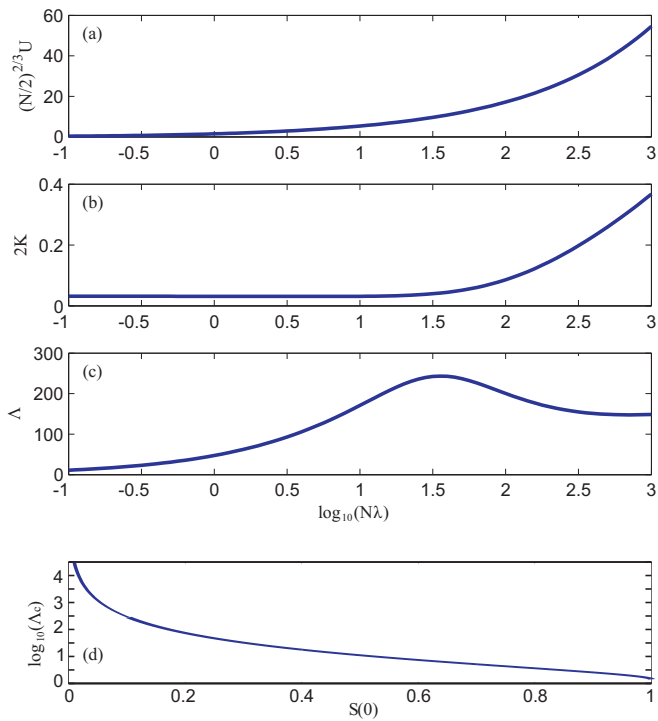}
\caption{(a) Strength of nonlinearity $(N/2)^{3/2}U$, (b) tunneling energy
$2{\cal K}$, and (c) their ratio $\Lambda$ (c) as functions of $N\lambda$ at
unitarity. (d) The critical $\Lambda_c$ as function of $S(0)$ for $\theta(0)=0$ [see Eq.~(\ref{Lambdac})].}
\label{Lambda}
\end{figure}

In Figs. \ref{Lambda} (a), (b), and (c) we illustrate the nonlinear 
strength $(N/2)^{2/3}U$, tunneling energy $2{\cal K}$ and their ratio 
$\Lambda$ as functions of $N\lambda$, respectively. As $N\lambda$ 
increases, the strength of the (repulsive) nonlinearity increases. As a 
result, the $\phi_1$ and $\phi_2$ become widened and enjoy more overlap. 
This leads to an increased tunneling energy. However, the ratio of the 
nonlinear strength and the tunneling energy $\Lambda$ does not have a 
monotonic behavior as $N\lambda$ is increased. As shown in 
Fig.~\ref{Lambda}(c), $\Lambda$ initially increases for small 
$N\lambda$, reaches a peak and then decreases. 
In a recent study,
Salasnich {\it et al.} \cite{newref}  used the local-density
approximation on top of quantum Monte-Carlo data of Ref.  \cite{the1} 
 to explore the phase diagrams and find  regimes
of Josephson tunneling and of dynamical self-trapping of a 3D
Fermi superfluid.
In the two-mode approach 
reported in Ref.~\cite{newref}, a constant tunneling energy is 
arbitrarily chosen for the whole crossover regime. This is an 
inappropriate oversimplification. In Fig.~\ref{Lambda}(d), we show the 
critical value $\Lambda_c$ as a function of the initial population 
imbalance $S(0)$. One can see that as $S(0)$ increases, $\Lambda_c$ 
decreases rapidly.

The fact that $\Lambda$ is bounded
from above even though the interaction strength can increase without
bound has important consequences. For example, for certain initial
conditions, self-trapping may only occur within an intermediate
range of nonlinearity. Both too small and too large a nonlinearity
will destroy self trapping. This statement is also true away from
the unitarity, even in the BEC limit. Furthermore, for a
sufficiently small
$S(0)$, $\Lambda$ may never exceed the corresponding $\Lambda_c$. When this is the case, the system will always stay in the Josephson oscillation regime. For example, according to Fig.~\ref{Lambda}(d),
$\Lambda_c \approx 300$ for $S(0)=0.1$. Fig.~\ref{Lambda}(c) shows that the system can
therefore never reach the self-trapping regime if $S(0)$ equals 0.1 or smaller.
This is consistent with our numerical results.

We want to remark that even though results obtained from the simple
two-mode model may provide significant qualitative insights, they are not expected to be accurate quantitatively. Particularly
for large nonlinearity, predictions from the two-mode model can
deviate greatly from the numerical results \cite{I2M,hanpu}. The
error mainly occurs in estimating the tunneling energy ${\cal K}$.
The two-mode equations (\ref{2modeA}) and (\ref{2modeB}) are
obtained by neglecting  many terms involving overlap integrals of
the mode functions $\phi_1$ and $\phi_2$, and hence in general
greatly underestimates the tunneling energy, particularly for large
nonlinearity when overlap between $\phi_1$ and $\phi_2$ can be
significant. Furthermore, the two-mode approximation itself becomes
questionable for large nonlinearity. When there is exchange of 
atoms
between the two wells, the mode functions will change accordingly
due to the modification of the nonlinear mean-field. Indeed, in our
numerical calculations to be presented below, we observe that the spatial wave function of
the system changes in time. In certain regimes, this change is significantly enough to invalidate the two mode model.

\section{Numerical results}

\label{III}

In this section we present an account of the numerical study of
self-trapping and Josephson oscillation in a double-well potential by solving
the full quasi-1D DF GP equation (\ref{eq6}) valid for a cigar-shaped SF. The
double-well potential is taken  as
\begin{equation}\label{well1}
U(x)=x^2/2+Ae^{-\kappa x^2}.
\end{equation}
We shall take the parameters $A$ and $\kappa$
of this double well
similar to the ones employed in
Ref. \cite{hanpu} in a study of self-trapping
with dipolar bosonic atoms.

To create an initial state with desired population imbalance for a given 
set of parameters $N\lambda$ and $a$, we search for the ground state of 
an asymmetric well comprised of an off-centered harmonic potential and 
the Gaussian barrier potential
\begin{eqnarray}\label{asy}
U'(x)=(x-x_0)^2/2+Ae^{-\kappa x^2} .  
\end{eqnarray}
The ground state of this asymmetric well is obtained by solving the time-independent version of the DF GP
equation (\ref{eq6}) using the imaginary time evolution method. The parameter
$x_0$ in (\ref{asy}) is chosen so that the population imbalance
\begin{equation}\label{eq7} S(t)=(N_1(t)-N_2(t))/N \,,
\end{equation} has a fixed pre-determined initial value $S(0)$. Here
$N_1(t)$ and $N_2(t)$ are the number of dimers in the first and the
second well of the double-well potential.
Experimentally, this is indeed the method used to generate the initial
population imbalance \cite{GatiJPB2007}. We have also considered
other forms of initial wave functions and found that the final
results are {{qualitatively insensitive to the specific forms
provided the
initial population imbalance $S(0)$ is kept fixed at a small
value.  However, at a quantitative level the results could be sensitive to the
form of the initial wave function. The sensitivity of the result to the
initial wave form increases as $S(0)$ is increased. Moreover,}} the
results are quite sensitive to the initial $S(0)$ employed.

Once the initial wave function is chosen, Equation (\ref{eq6}) is
solved numerically after discretization with the Crank-Nicolson
scheme \cite{MA,MA2}
employing space and time steps 0.025 and 0.0002,
respectively, using real-time propagation with the FORTRAN programs
provided in Ref. \cite{MA}. The results are also independently
confirmed using a MATLAB code based on the split fast Fourier transform method.

\begin{figure}
\includegraphics[width=\linewidth]{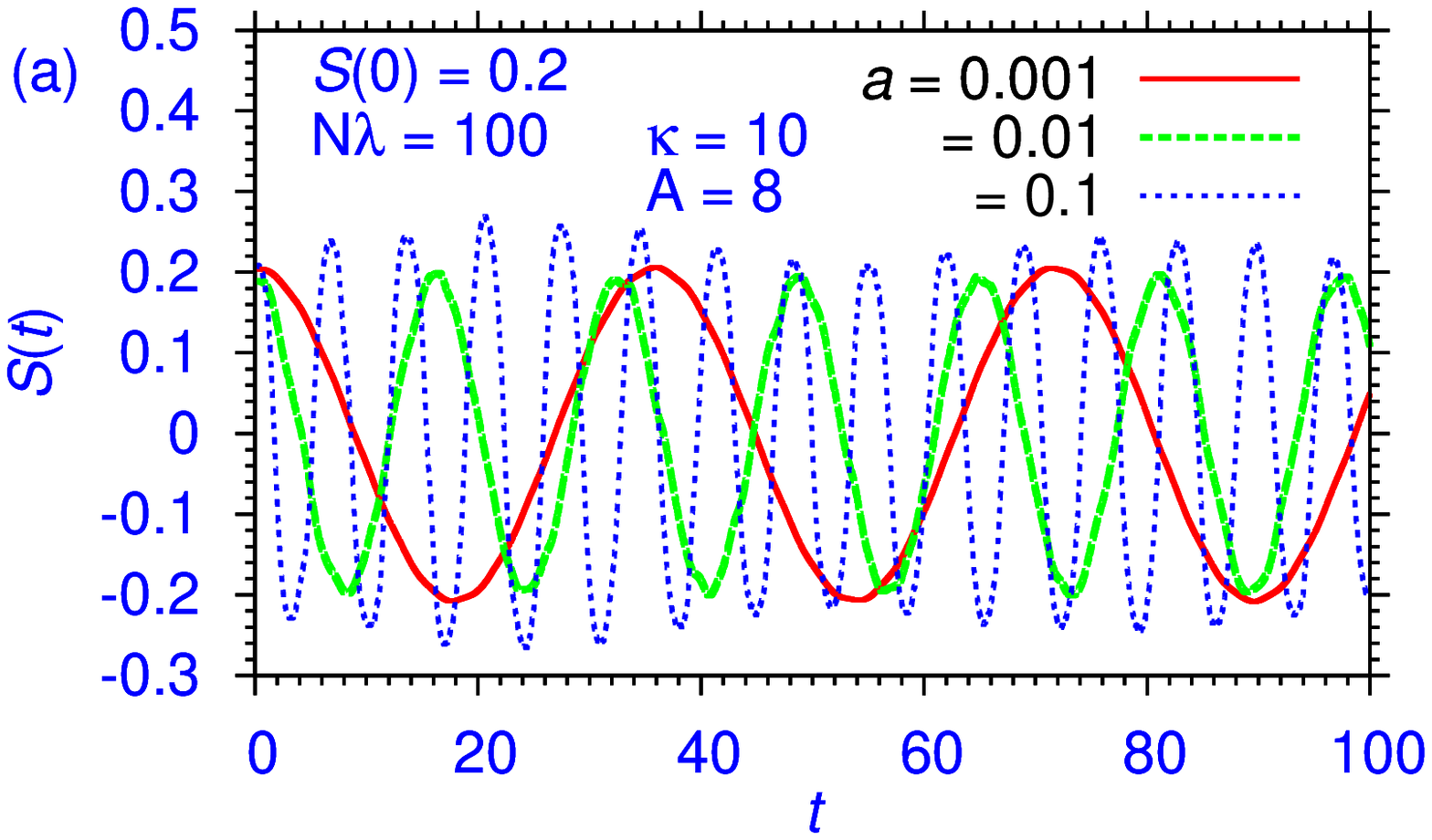}
\includegraphics[width=\linewidth]{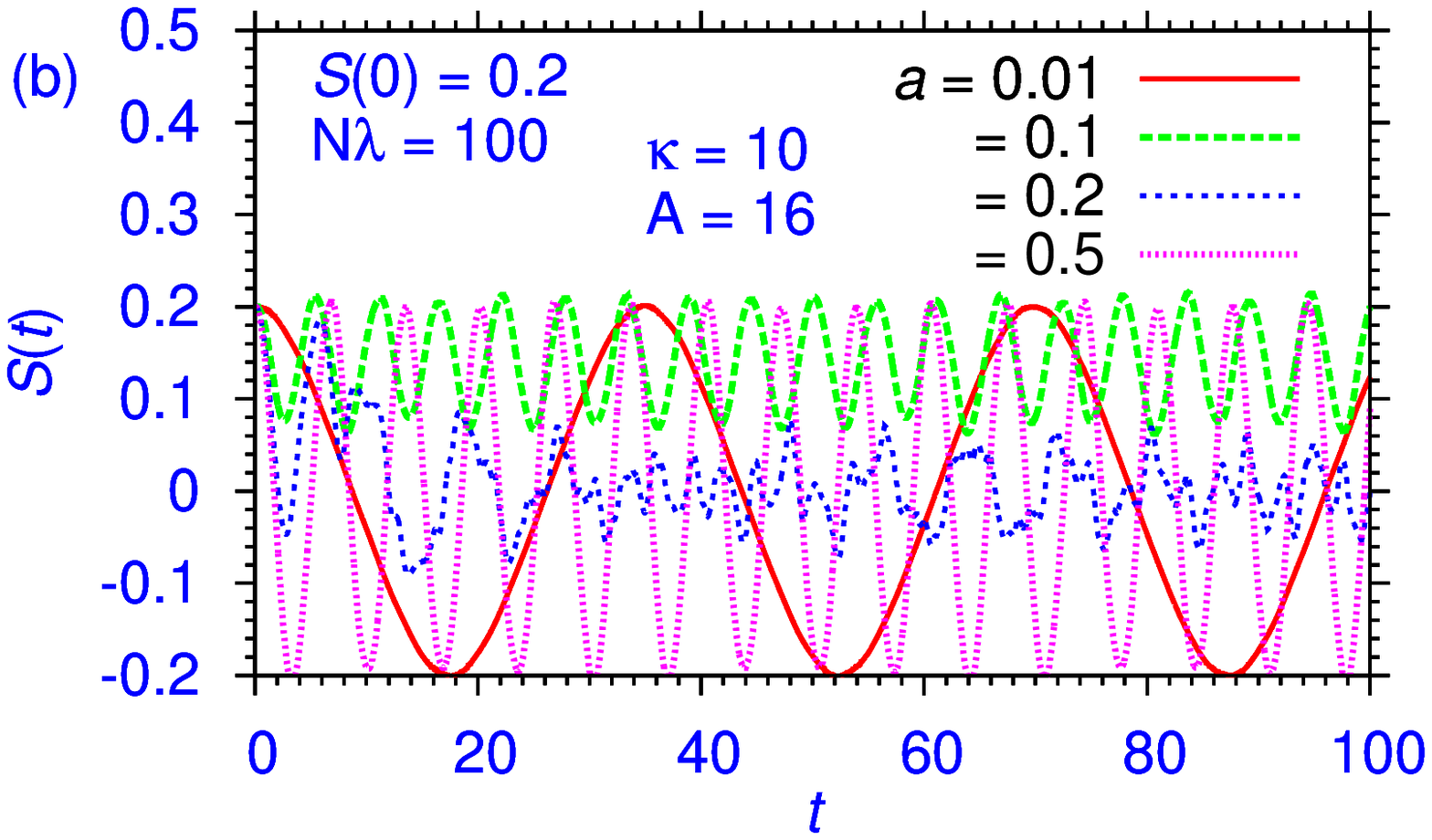}
\includegraphics[width=\linewidth]{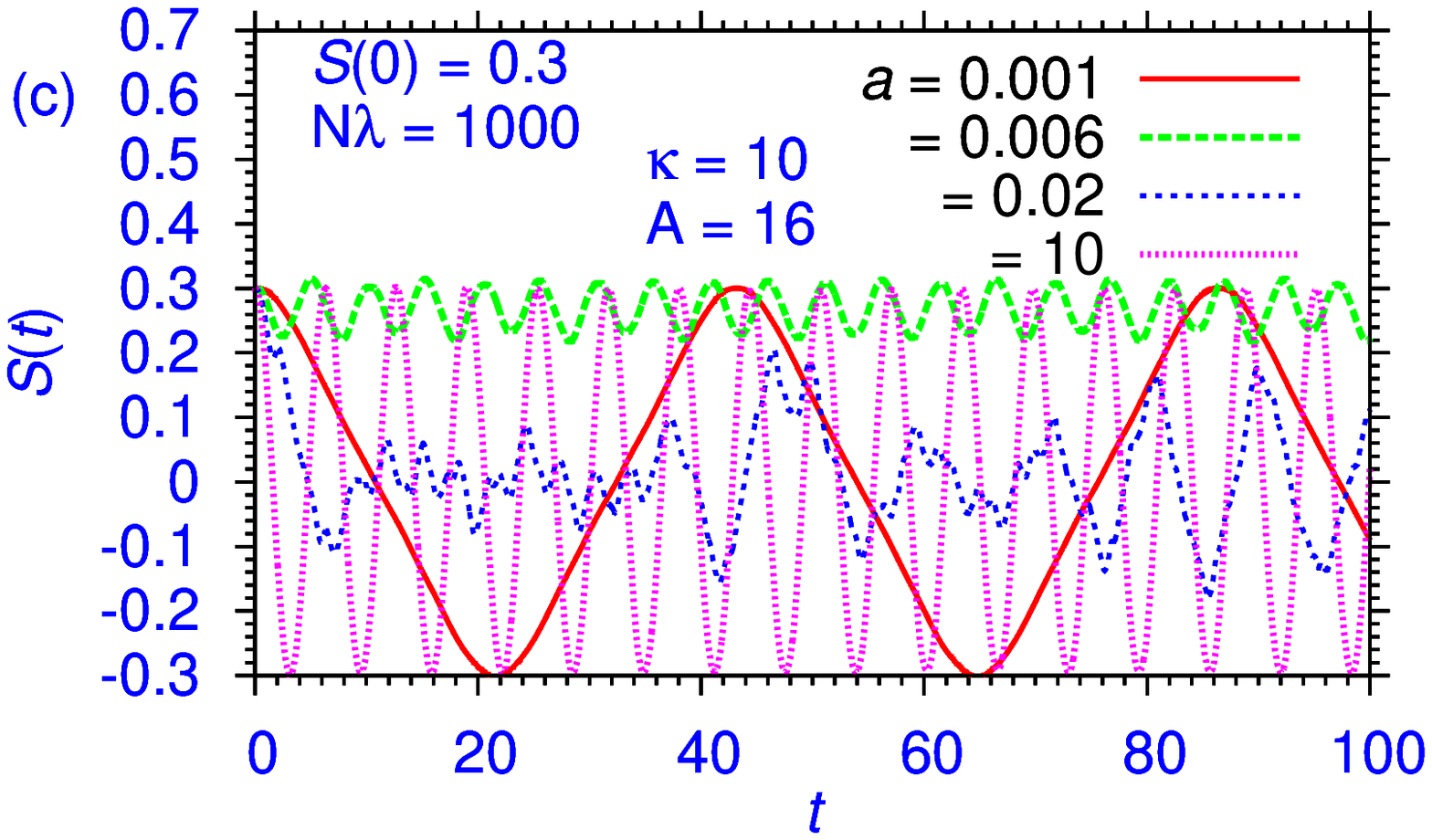}
\includegraphics[width=\linewidth]{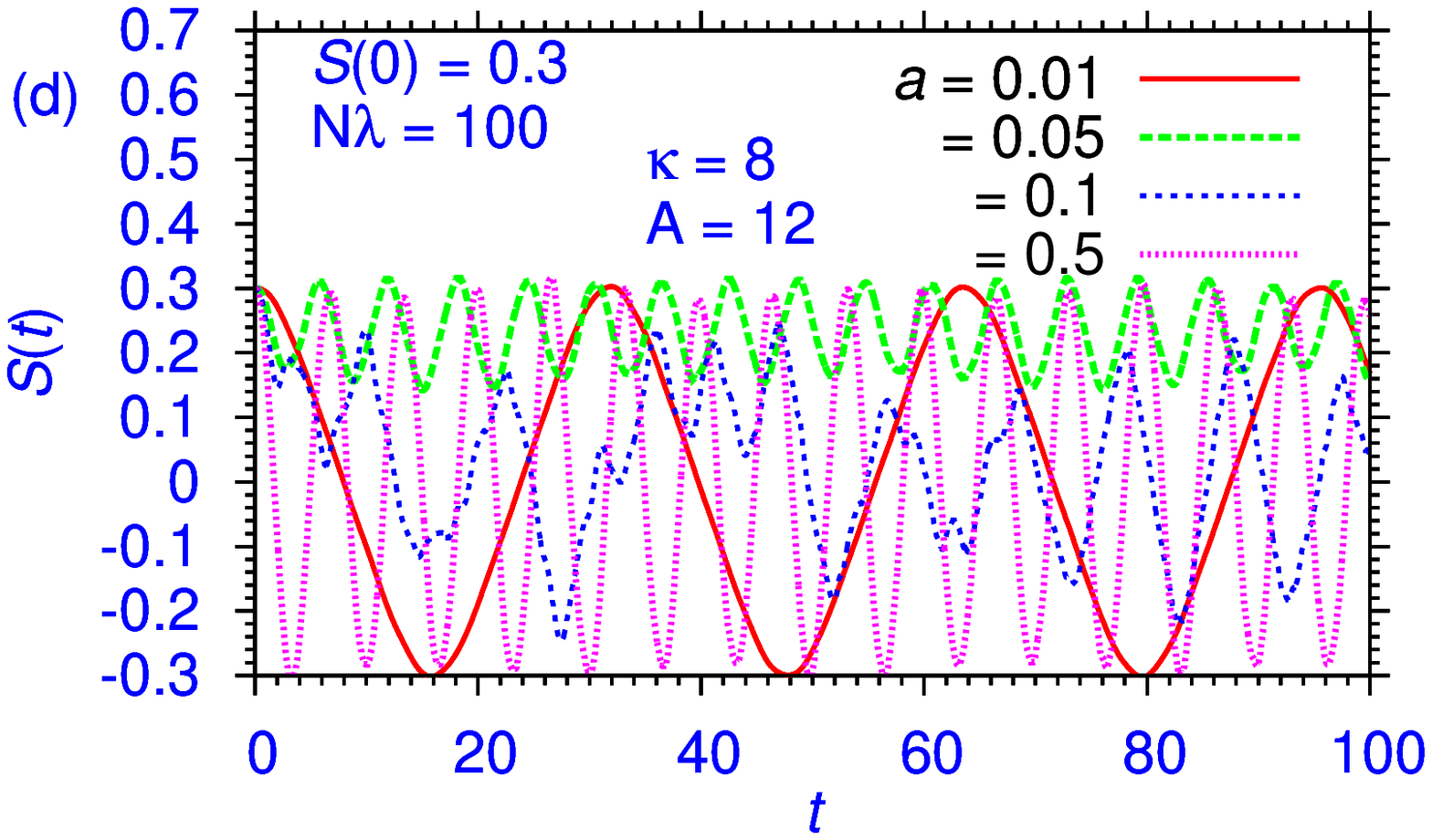}
\caption{(Color online) Population imbalance $S(t)$ vs. $t$ dynamics
with 
(a) $N\lambda =100, S(0)=0.2, \kappa=10, A=8$, 
(b) $N\lambda =100, S(0)=0.2, \kappa=10, A=16$, 
(c) $N\lambda =1000, S(0)=0.3, \kappa=10, A=16$, 
(d) $N\lambda =100, S(0)=0.3, \kappa=8, A=12$, 
for dimer-dimer scattering length $a$
varying over the BEC-unitarity crossover. }
\label{F3}
\end{figure}

Now we present results of dynamical evolution of a Fermi SF, 
where we
take $\xi =13.37$ in Eqs. (\ref{eq6}) and (\ref{eq6x}).
The numerical study of self-trapping and Josephson oscillation with
the Fermi SF of dimers along the BEC-unitarity crossover reveals
interesting features. 
{ To start the investigation of self-trapping we fix the trap 
parameters $A$ and $\kappa$ in Eqs. (\ref{well1}) and (\ref{asy}) at 
nontrivial values (a not too small value of $A$ and a not too large 
$\kappa$), which permit smooth and free Josephson oscillation in the BEC 
limit ($a=0$). Note that a very small value of $A$ and a very large 
value of $\kappa$ tend to reduce the double well (\ref{well1}) to a 
single well where there cannot be any self trapping and Josephson 
oscillation should appear for all values of $a$ and $N$. In order to 
have self trapping, $A$ cannot be too small and $\kappa$ cannot be too 
large. 
This is illustrated in Fig. \ref{F3} (a) for $A=8, \kappa=10, 
N\lambda =100, S(0)=0.2$ and for $a=0.1,0.01,$ and 0.001 where we plot 
$S(t)$ vs. $t$. 
There is no self trapping for a very small value of 
$S(0) (=0.1)$.
The quantity $S(t)$ is experimentally measurable and 
$S(t)$ vs. $t$ dynamics provides information about self trapping and 
Josephson oscillation. From Fig. \ref{F3} (a) we find that there is 
Josephson oscillation for all values of $a$ and there is no sign of self 
trapping. (A nonzero time average $\langle S(t) \rangle$ ensures self 
trapping.)  But a completely new scenario emerges as $A$ is increased to 
16 from 8, as can be seen from Fig. \ref{F3} (b) where we show the 
$S(t)$ vs. $t$ dynamics for $a=0.01, 0.1,0.2$ and 0.5. The plots for 
$a=0.01$ and 0.5 of Fig. \ref{F3} (b) are quite similar to the plots 
for $a=0.001$ and 0.1 of Fig. \ref{F3}  (a) illustrating regular 
(periodic)
Josephson oscillation
with no sign of self trapping.  But, for intermediate values 0.1 and 0.2 
of $a$, self trapping and irregular (non-periodic) oscillation can 
be seen in Fig. \ref{F3} (b). In  Figs. \ref{F3} (c) and (d) 
we illustrate two more cases of $S(t)$ vs. $t$ dynamics 
with a different value of $S(0) (=0.3)$
and 
different $N\lambda$ and trap parameters, respectively, where one can clearly 
find self trapping.
}

In the following,
we discuss in detail the results for three initial
population imbalance $S(0)=0.1$, 0.2 and 0.3, which are representative 
for a general case.

{\it Population imbalance} $S(0)=0.1:$
For this relatively small initial population imbalance, we found that for any
values of $N\lambda$ and $a$, the system is always in the Josephson regime:
the population imbalance $S(t)$ oscillates sinusoidally between
$S(0)=-0.1$ and $S(0)=0.1$. The system never exhibits self-trapping.
The frequency of oscillation increases as the strength of
nonlinearity increases.
Note that the strength of nonlinearity is increased
by increasing either $N\lambda$ or $a$.
However, the nonlinear interaction among
dimers saturates as scattering
length $a\to \infty$ at unitarity, it increases indefinitely with $N\lambda$. This result is consistent with our previous discussion of the two-mode model: For a sufficiently small $S(0)$, the required critical value of $\Lambda_c$ for self-trapping cannot be achieved by increasing the strength of the nonlinearity and the system stays in the Josephson regime for all values of $N\lambda$ and $a$.

{\it Population imbalance} $S(0)=0.2:$
The results for the $ S(t)$ vs. $t$ dynamics for
this initial population imbalance is
illustrated in Figs.~\ref{F3} (a) and  (b) for fixed $N\lambda=100$ and various
values of $a$ and two different traps as described earlier. 
In Fig. \ref{F3} (b),
for a small scattering length of $a=0.01$ (solid line),
Josephson oscillation is observed. When $a$ is increased to 0.1 (dashed
line), self-trapping is clearly seen --- $S(t)$ does not deviate from
$S(0)$ too much and never crosses zero. With further increase of $a$ to a
slightly larger value of 0.2 (dotted line), self-trapping is destroyed
and $S(t)$ exhibits irregular oscillations around zero. Accompanied with
this irregular population oscillation, the density profile
$|\psi(x,t)|^2$ also develops complex and irregular structures.
Remarkably, upon further increase of  $a$, as the dot-dashed curve for
$a=0.5$ shows, regular oscillation returns and the population dynamics
once again exhibits sinusoidal Josephson oscillations.

{\it Population imbalance} $S(0)=0.3:$
Finally, let us discuss this relatively large initial
population imbalance. If we use $N\lambda =100$ as we did above for
$S(0)=0.2$, when $a$ is increased from zero to $\infty$, the system
sequentially makes transitions from Josephson, to self-trapping and
finally to irregular oscillation regimes. {{The Josephson
oscillation is never recovered for $N\lambda =100$} for very large values of
scattering length $a$ (results not shown here). In Fig.~\ref{F3} (c) we plot
the results for $ S(t)$ vs. $t$ dynamics for   
$N\lambda=1000$. In this case}},
in
addition to the three regimes just mentioned, for a  sufficiently large
$a {{ (= 10)}}$, Josephson oscillation is restored, just as in
the case of $S(0)=0.2$ and $N\lambda=100$
discussed above.
{In Fig.  Fig.~\ref{F3} (d) we show another example of $ S(t)$ vs. $t$ dynamics
for a different trap, which is quite similar to that in Fig.~\ref{F3} (c). 
We also did some calculation with larger $S(0)$ where a similar panorama 
emerges and we do not report the details here.}

\begin{figure}
\includegraphics[width=\linewidth]{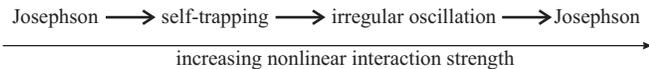}
\caption{Different dynamical regimes of Fermi super-fluid in the BEC-unitary crossover.}
\label{sum}
\end{figure}

To summarize the { general characteristics of the}
population dynamics, we find that for any given initial 
population imbalance and for either sufficiently small or sufficiently 
large nonlinear interaction strength, the system is in the Josephson 
oscillation regime. For intermediate interaction strength, the system 
can make transition to self-trappping and irregular oscillation regimes 
as schematically shown in Fig.~\ref{sum}. The critical interaction 
strength at which the system makes the transition to self-trapping is 
sensitive to the initial population imbalance and increases sharply as 
$S(0)$ increases.  (It is also sensitive to the parameters for the 
Gaussian barrier that creates the double-well potential.) 
It is possible that for a sufficiently small $S(0)$, 
the system always stays in the Josephson regime. The restoration of the 
Josephson oscillation at large interaction strength may seem surprising 
at first sight. However, one can understand it in the following 
intuitive way. For a sufficiently large interaction strength, the chemical 
potential is large and the effect of the Gaussian barrier becomes 
relatively unimportant. The wave functions on opposite sides of the 
barrier have sufficient overlap and hence the cloud tunnel back and 
forth without difficulty.

\begin{figure}
\begin{center}
\includegraphics[width=\linewidth]{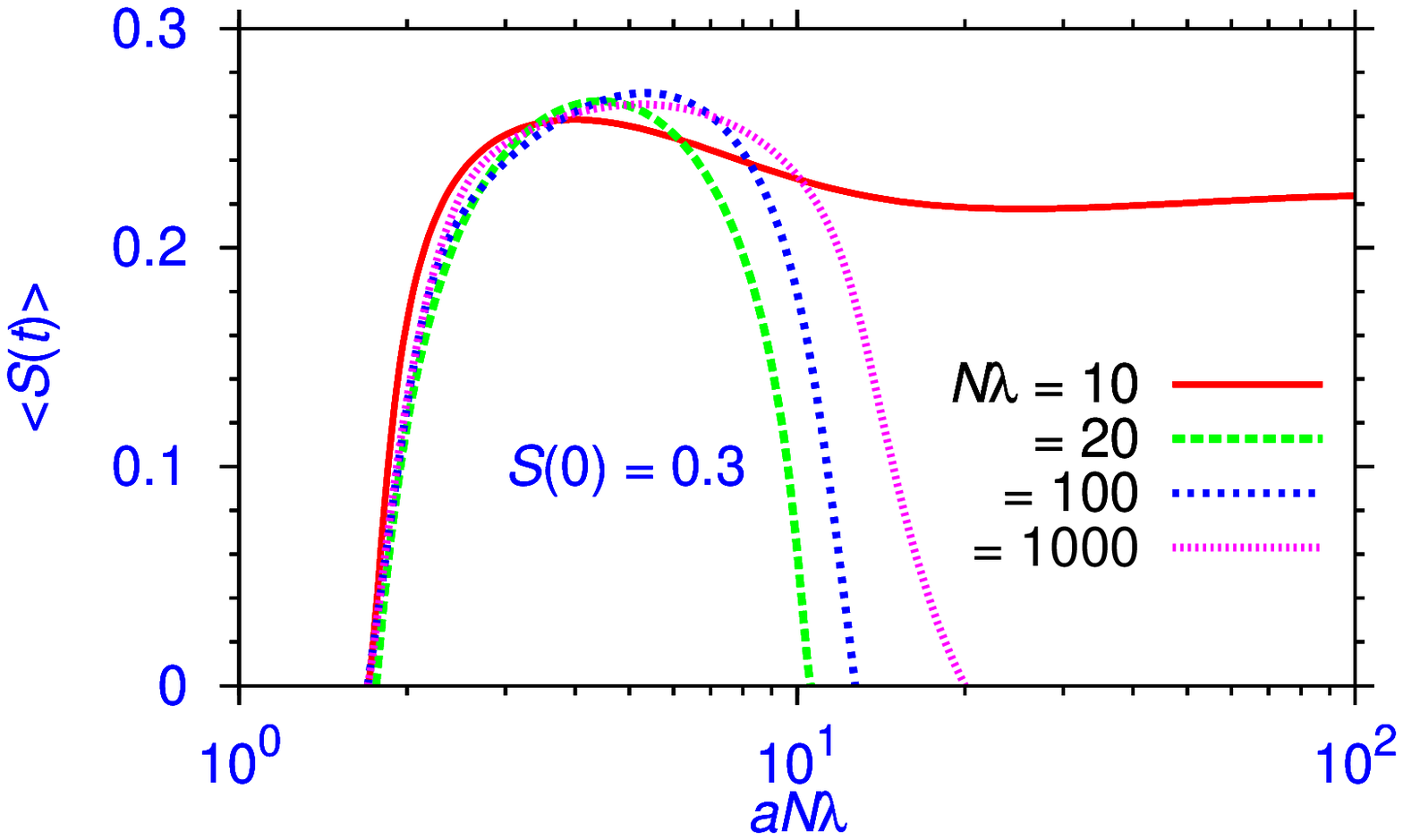}
(a)
\includegraphics[width=\linewidth]{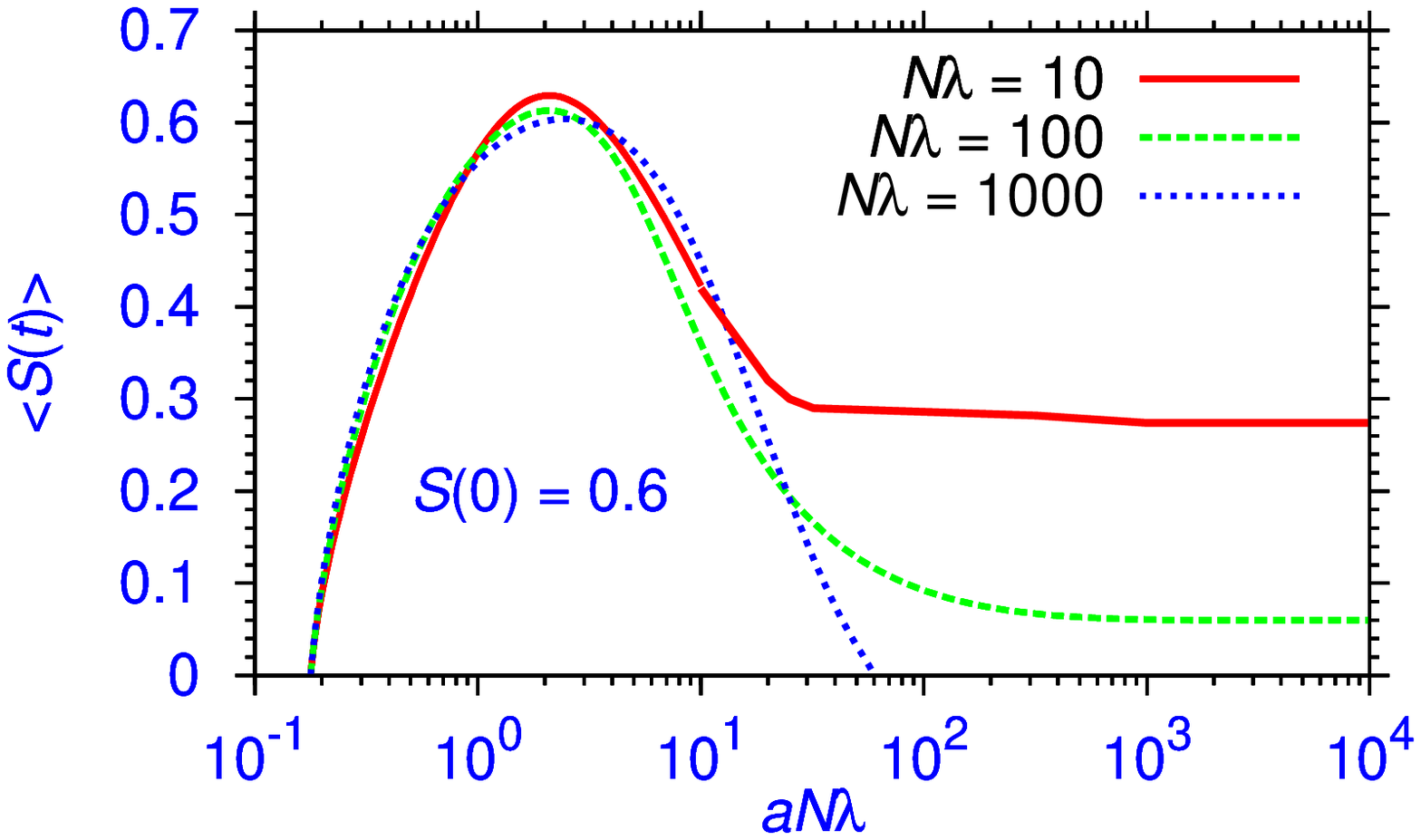}
(b)
\end{center}

\caption{(Color online) Time-averaged population imbalance
$\langle S(t) \rangle$ vs.
nonlinearity  $aN\lambda$ for different $N\lambda$ for trap parameters 
$A=16, \kappa=10$ and (a) $S(0)=0.3$ and (b) $S(0)=0.6$. Different
curves are generated by varying the scattering length $a$ across the
BEC-unitarity crossover for  $S(0)=0.3$ and 0.6, respectively, 
based on initial
wave form (\ref{asy}).
}
\label{F5}
\end{figure}

The appearance of self-trapping is best illustrated through a study
of the time-averaged population imbalance $\langle S(t) \rangle$ vs.
nonlinearity $aN\lambda$ and we do that next. In Fig. \ref{F5} (a) we
plot $\langle S(t) \rangle$ vs. $aN\lambda$ by varying the
scattering length from 0 to $\infty$ for a fixed $N\lambda$ with
trap parameters $A=16$ and $\kappa=10$. 
The initial population imbalance is chosen as $S(0)=0.3$.
In Fig.~\ref{F5} (a), with the increase of $aN\lambda$,
self-trapping appears for $aN\lambda$ slightly greater than unity.
With further increase of $aN\lambda$, self-trapping increases with
an increase of $\langle S(t) \rangle$. For $N\lambda =10$, self
trapping never disappears and continues even at unitarity. However,
beyond $aN\lambda\approx 5$ self-trapping decreases with the
increase of $aN\lambda$ for $N\lambda=20, 100$ and 1000. For larger
$N\lambda,$ $\langle S(t) \rangle$ eventually goes to zero as the
nonlinear repulsion becomes too large to maintain all dimers in a
single trap, except for $N\lambda =10$. 
{ In Fig.~\ref{F5} (b) we exhibit a 
similar plot for $S(0)=0.6$
with 
trap parameters $A=16$ and $\kappa=10$. 
 }

Finally, to check the validity of the quasi-1D approximation, we 
performed full 3D numerical simulations based on Eqs.~(\ref{eq3}) and 
(\ref{eq32}). The quasi-1D approximation should be valid when $\mu \ll 
\lambda \hbar \omega$. We have chosen different sets of parameters, some 
of which satisfy and the rest violate the quasi-1D condition. For 
parameters such that the quasi-1D condition is satisfied, we indeed find 
that our 3D numerical results are nearly identical to the 1D results 
presented here. For parameters that the quasi-1D condition is violated, 
the 3D results show deviations from the 1D ones. However, the 
qualitative features (i.e., the dependence of different dynamical 
regimes on the initial population imbalance and the strength of 
nonlinearity) presented in Figs.~\ref{F3} and \ref{F5} remain valid. 
{Specifically, we verified that the results reported in Figs. \ref{F3} 
remain essentially  valid in the 3D model. This assures us of the reliability 
of the present study in 1D. }

\section{Summary and Conclusion}
\label{V}

To summarize, we have studied the dynamical properties of a Fermi SF
confined in a double-well potential in the BEC-unitary crossover
regime. To this purpose, we have developed a nonlinear
Sch\"{o}dinger equation valid in the whole regime based on a density
functional approach and on the equations of state from quantum Monte
Carlo calculations. This equation is equivalent to the hydrodynamic
equations with the quantum pressure term included. In the BEC side
of the crossover, it describes accurately the equilibrium and
low-energy dynamical properties of the Fermi SF. In particular,
Josephson effect has been investigated using this method \cite{sala}
and the results have been shown to agree with those obtained from
the microscopic approach by solving the Bogoliubov-de Gennes
equations \cite{bdg}. Compared with the latter, the great advantage
of the current approach is its mathematical simplicity.

We have
identified three dynamical regimes of the system: the Josephson
regime, the self-trapping regime and the irregular oscillation
regime. For a given initial population imbalance, these regimes are
accessed according to the strength of nonlinearity as schematically shown in Fig.~\ref{sum}. The Josephson regime is always reached at either sufficiently small or sufficiently large interaction strength. 
For a small initial population imbalance, Josephson regime may be the only regime that the system can have access to. Note
that the strength of nonlinearity can be increased by either
increasing the number of dimers $N$ or the scattering length $a$.
However, it saturates as $a$ tends to infinity while no saturation
occurs for large $N$.

The quasi-1D model $-$ Eqs. (\ref{eq6}) and (\ref{eq6x})  $-$
presented and used in the study of dynamical evolution of a Fermi SF in the BEC-unitarity crossover in this paper
is also valid for an atomic BEC with a slightly
modified value for the parameter
$\xi$. Hence the present results for self-trapping of a Fermi SF in a
double-well
potential are also applicable for a repulsive atomic BEC when the atomic scattering length varies from 0 to $\infty$.
However, in this case there could be practical difficulty with three-body loss
in the experimental realization of the system for large scattering length.

We have also developed a simple analytical two-mode model, analogous
to the much studied system of a BEC in a double-well potential. We show that the properties of the system can be described by a classical Hamiltonian with population imbalance and relative phase as a pair of conjugate variables. The great advantage of the two-mode model is its simplicity which makes analytical studies possible. The key parameters that characterize the two-mode model are the strength of nonlinearity and the tunneling energy. We calculated these parameters using the spatial mode function obtained by numerically solving the full time-independent nonlinear Schr\"{o}dionger equation. From this calculation we show that the ratio of the interaction strength and the tunneling rate cannot increase indefinitely when the interaction strength increases. This explains the numerical observation that for sufficiently small initial population imbalance, the system may always stay in the Josephson regime. However, care must be taken when making quantitative comparisons with numerical results. In particular, for strong nonlinearity, the two-mode model can be even qualitative incorrect. For example,
this model predicts the existence of the Josephson and the
self-trapping regime, but not the irregular oscillation regime 
found
in the numerical calculation, which occurs at 
relatively large nonlinearity and
lies in the regime where the two-mode model is no longer
valid.

\acknowledgments

FAPESP and CNPq (Brazil) provided partial support. H.P. acknowledges support from NSF and the Robert A. Welch Foundation (Grant No. C-1669).

\end{document}